\documentclass[a4paper,11pt]{article}

\usepackage{pos}
\usepackage[english]{babel} 
\usepackage[utf8]{inputenc}
\usepackage{amsthm} 
\usepackage{mathtools}
\usepackage{pgfplots}
\usepackage{tensor} 
\usepackage{physics}
\usepackage{bm}

\newcommand{\cov}[2]{{\rm cov}[#1,#2]}

\newcommand{\Corr}{\mathrm{Corr}}

\title{A variance reduction technique for hadronic correlators with partially twisted boundary conditions}
\ShortTitle{Variance reduction for partially twisted boundary conditions}

\author[a,b]{Nils Asmussen}
\author*[a,b]{Alessandro Barone}
\author[a,b,c]{Andreas J\"uttner}

\affiliation[a]{School of Physics and Astronomy, University of Southampton, Southampton SO17 1BJ, UK}
\affiliation[b]{STAG Research Center, University of Southampton, Southampton SO17 1BJ, UK}
\affiliation[c]{CERN, Theoretical Physics Department, Geneva, Switzerland}

\emailAdd{a.barone@soton.ac.uk}

\abstract{
Partially twisted boundary conditions are widely used for improving the momentum resolution in lattice computations of hadronic correlation functions. The method is however expensive since every additional twist requires computing additional propagators.
We propose a novel variance reduction technique that exploits statistical correlations to reduce the overall cost for computing correlators with additional twist angles. We explain and demonstrate the method for meson 2pt and 3pt functions.
}

\FullConference{%
 The 38th International Symposium on Lattice Field Theory, LATTICE2021
  26th-30th July, 2021
  Zoom/Gather@Massachusetts Institute of Technology
}

\begin{document}
\maketitle

\section{Introduction}

In a finite size cubic volume with periodic boundary conditions, the hadronic momenta $\bm{p}$ are quantized in each spatial direction $i=x,y,z$ as $p_{i}=\frac{2\pi}{L}n_i$, where 
$L$ is the spatial extent of the lattice and $n_i$ takes integer values.
This implies that accessible values of $\bm{p}$ in a lattice simulation are limited. 
In view of lattice QCD simulations for SM phenomenology this often constitutes a limiting factor, e.g. when a fine momentum resolution in the computation of hadronic form factors is required.
A way to overcome this limitation is to impose partially twisted boundary conditions \cite{SachVill, deDiv} on a valence quark field $q$ as
\begin{align}
 q(x_{i}+L) = e^{i\theta_i}q(x_{i})\, , \quad 0 \leq \theta_i \leq 2\pi \, , \quad i=1,2,3 \, ,
\end{align}
where $\theta_i$ is the twisting angle. 
For a meson with one quark twisted with angles $\bm{\theta}=(\theta_1, \theta_2, \theta_3)$ and the other one untwisted, the
twist has the effect of shifting the corresponding meson momentum as
\[
 p_{i} = \frac{2\pi}{L}n_i +\frac{\theta_i}{L} \, ,
\]
thereby extending the accessible range of momenta \cite{FlynnJutt_Numerical, FlynnJutt}. 

\section{General idea}

In this work we propose a novel technique to reduce the overall cost for computing correlators with additional twist angles.
From a computational point of view, twisting comes with an extra cost since every momentum requires computing additional quark propagators.
The goal of this work is to induce momentum into hadrons through twisted boundary conditions with less computational effort.

We assume that all the untwisted correlation functions $C_{0}(t)$ have already been computed on a given ensemble of gauge configurations.
The idea is to exploit these correlators in order to reduce the resources needed to obtain twisted correlation functions  from propagators evaluated on $N_{\rm src}$ source planes with the same precision. We construct a new twisted correlator $C_{\theta}^{\alpha}(t)$ as
\begin{align}
C_\theta^{\alpha}(t) =\tilde C_\theta (t) + \alpha(t)(C_0 (t)-\tilde C_0 (t))\,,
\label{eq:C_alpha}
\end{align}
where the expressions with tilde are obtained with a reduced number of source planes, and $\alpha(t)$ is an optimisation parameter, which in general depends on the time $t$. In particular, for a generic correlator $C(t)$ we use the notation
\begin{align*}
  C(t) = \frac{1}{N_{\rm src}}\sum_{i=1}^{N_{\rm src}} \langle C_i (t)\rangle \, , \quad
  \tilde{C}(t) = \frac{1}{n_{\rm src}} \sum_{i=1}^{n_{\rm src}} \langle C_i (t) \rangle \, ,  \quad n_{\rm src}<N_{\rm src} \, ,
\end{align*}
where the sum is running over the source planes and $\langle \cdot \rangle$ denotes the ensemble expectation value.
Expressions with $\alpha$ as superscript refer to the improved case built with the new correlators \eqref{eq:C_alpha}. We will often omit the $t$ depence in $\alpha$ for simplicity.
The case where $\alpha=0$ corresponds to the standard twisted scenario with a reduced number of sources.\\
Assuming the measurements on a given time slice to be normally distributed, the expression for the variance 
$(\sigma_{\theta}^\alpha)^2=\langle (\Delta C^{\alpha}_{\theta})^2 \rangle$ is
\begin{align*}
(\sigma_{\theta}^\alpha)^2=&
\alpha^2(\sigma_{0}^2+\tilde{\sigma}_{0}^2
- 2\cov{C_0}{\tilde C_0})+
2\alpha(\cov{ {C_0}}{{\tilde C_\theta}}-\cov{\tilde C_0}{\tilde C_\theta})
+\tilde{\sigma}_{\theta}^2 \, ,
\end{align*}
where $\cov{\cdot}{\cdot}$ is the corresponding covariance for that time slice.
The extrema under variation of $\alpha $ are determined through
\begin{align*}
\left. \pdv{\alpha} (\sigma_\theta^{\alpha})^2 \right|_{\alpha_{min}}=&
2\alpha_{min} (\sigma_{0}^2+ \tilde{\sigma_{0}}^2
- 2\cov{C_0}{\tilde C_0})
+ 2(\cov{C_0}{\tilde C_\theta}-\cov{\tilde C_0}{\tilde C_\theta}) =0\,.
\end{align*} 
The value 
\begin{align}
\alpha_{min} =\frac{\cov{\tilde C_0}{\tilde C_\theta}-\cov{C_0}{\tilde C_\theta}}{\sigma_{0}^2+ \tilde{\sigma}_{0}^2-2\cov{C_0}{\tilde C_0}}
\label{eq:alphamin}
\end{align}
is therefore expected to optimise the variance, which now reads
\begin{align}
 ( \sigma_\theta^{\alpha})^2_{min} =  \tilde{\sigma}_{\theta}^2 - \frac{\left(\cov{\tilde C_0}{\tilde C_\theta}-\cov{C_0}{\tilde C_\theta}\right)^2 }{\sigma_{0}^2 + \tilde{\sigma}_{0}^2 - 2 \cov{C_0}{\tilde{C}_0} } \, .
\end{align}
From this last expression we see that the optimized variance for the new correlator $C_{\theta}^{\alpha}$ is always smaller than the variance of the initial correlator $\tilde{C}_{\theta}$.

\section{Numerical setup}

We tested the above construction on a $24^3\times 64$ lattice using RBC/UKQCD DWF C1 ensemble \cite{RBC-UKQCD:2008mhs}, which has a pion mass of $m_{\pi}\simeq 330 \,\text{MeV}$ and a lattice spacing $a\simeq 0.11\,\text{fm}$. In particular, we considered 2pt pion ($C_{\pi}(t)$) and kaon ($C_{K}(t)$) correlators. In addition, we also computed $K\rightarrow\pi$ 3pt correlation functions $C_{\mu, K\pi}(t)$ 
(see \cite{Hadronic, kaon} for definitions), where $\mu$ labels the component of the light-strange current, 
as required for the computation of the $K\rightarrow\pi$ semileptonic form factors \cite{Hadronic, kaon}.
For the 3pt functions we used a source-sink separation of $t/a=20$.
The twist has been applied on the light quark in all the three spatial directions; in particular, we set
\begin{align}
 \begin{dcases}
  \bm{\theta_1} &= 0.1\times(2\pi, 2\pi, 2\pi) \, , \\
  \bm{\theta_2} &= 0.2\times(2\pi, 2\pi, 2\pi) \, , \\
  \bm{\theta_3} &= 0.3\times(2\pi, 2\pi, 2\pi) \, , \\
  \bm{\theta_4} &= 0.4\times(2\pi, 2\pi, 2\pi) \, , \\
 \end{dcases}
 \label{eq:twists}
\end{align}
%
%
which correspond to a total induced momentum of
\begin{align}
 \begin{dcases}
  |\bm{p_1}| &= 0.1\sqrt{3} \, \frac{2\pi}{L} \simeq 0.17\,\frac{2\pi}{L} \, , \\
  |\bm{p_2}| &= 0.2\sqrt{3} \, \frac{2\pi}{L} \simeq 0.35\,\frac{2\pi}{L} \, , \\
  |\bm{p_3}| &= 0.3\sqrt{3} \, \frac{2\pi}{L} \simeq 0.52\,\frac{2\pi}{L} \, , \\
  |\bm{p_4}| &= 0.4\sqrt{3} \, \frac{2\pi}{L} \simeq 0.69\,\frac{2\pi}{L} \, . \\
 \end{dcases}
\end{align}
%
%
In the next sections, we will indicate the twisting angle simply as $\theta=0.1, 0.2, 0.3, 0.4$ for simplicity, with obvious reference to the actual twist in \eqref{eq:twists}. \\
We generated a total of $N_{\rm src}=32$ sources on $39$ gauge configurations, and we considered a subset of $n_{\rm src}=4,8,16$ sources to test the method. The simulations have been carried out on the DiRAC Extreme Scaling service at the University of Edinburgh using the Grid\cite{Grid, Grid2} and Hadrons\cite{Hadrons} software packages.

\section{Results}

In this section we present some of the results obtained. We start with the results for basic quantities such as 2pt and 3pt correlators. We then present some further optimization on the observables built from these correlators, i.e. the $K\rightarrow\pi$ semileptonic form factors, which involve a combination of the former correlators.

\subsection{Correlators}

\begin{figure}[h]
 \centering
 \includegraphics[scale=0.4]{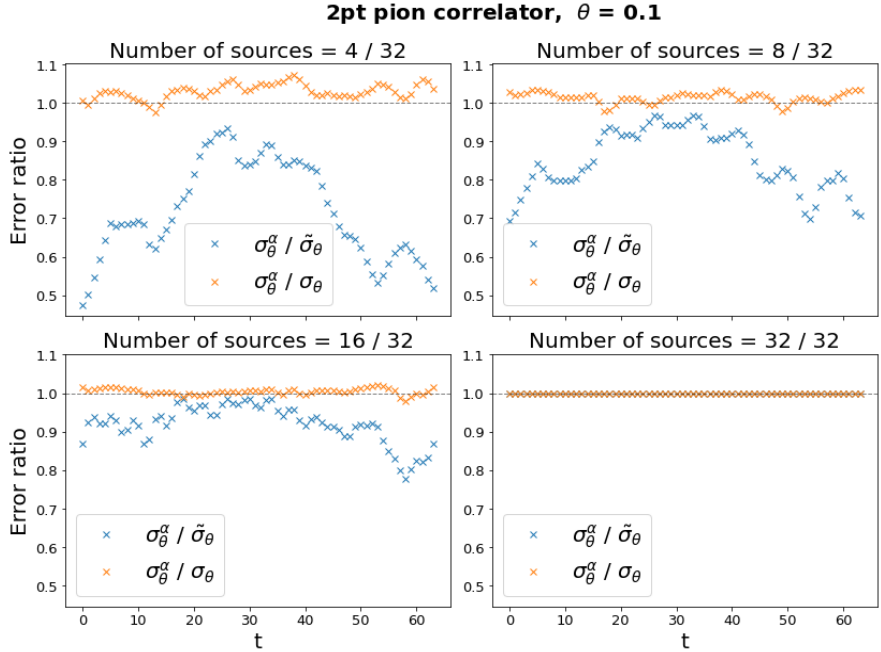}
 \caption{Improvement for a pion correlator with twist $\bm{\theta}=0.1(2\pi, 2\pi, 2\pi)$. The blue points correspond to the ratio of the improved correlator and the non-improved correlator with reduced statistics, whereas the orange points correspond to the ratio of the improved correlator and the non-improved correlator with full statistics.}
 \label{fig:pion_2pt}
\end{figure}

In figure \eqref{fig:pion_2pt} we show the first results for a pion correlator using the smallest twist $\theta=0.1$.
In the plot, we compare the error $\sigma_{\theta}^{\alpha}$ of the improved correlator $C_{\theta}^{\alpha}$ ($\alpha=\alpha_{min}$) with the standard correlator ($\alpha=0$) with reduced (blue) and full (orange) statistics. The plots show that already for 4 or 8 sources out of the total 32 there is a quite significant improvement. Indeed, the closer the orange points are to $1$, the closer the error of the improved correlator is to the full-statistics one. The blue crosses keep track of how much better we do with respect to the case $\alpha=0$ at reduced number of sources. In short, the method is successful as long as the blue dots move towards 0 and the orange dots move towards 1 (or lower). Looking at figure \eqref{fig:pion_2pt}, a good compromise would be the case with $8$ sources: at $1/4$ of the total computational cost we manage to get an error which is only few percent worse than the one with full statistic, but around $20\%$ better than what we would get without improving the correlator in the range where the ground-state dominance starts setting in. 

\begin{figure}[h]
 \centering
 \includegraphics[scale=0.38]{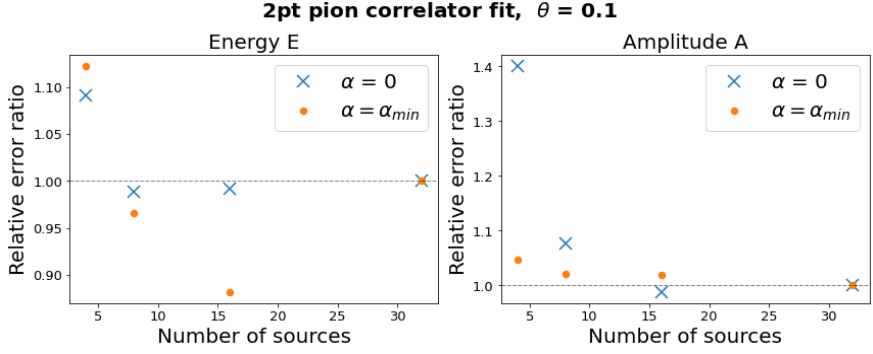}
 \caption{Relative error ratio $\tilde{\sigma}_{rel}/\sigma_{rel}$ as a function of the number of sources for improved (orange dots) and not improved (blue crosses) correlators.}
 \label{fig:fitpion_2pt}
\end{figure}

Let us now take a look at the physical quantities one can extract from this correlator, namely mass and amplitude. We fit the correlator using the ansatz
\begin{align}
   C_{\pi}(t) = 2Ae^{-E\frac{T}{2}}\cosh\left[E\left(\frac{T}{2}-t\right) \right] \, .
 \label{eq:fit}
\end{align}
The results are shown in figure \eqref{fig:fitpion_2pt}. There, the vertical axis represents the ratio of the relative errors $\tilde{\sigma}_{rel}/ \sigma_{rel}$ obtained from the fits as a function of the number of sources on the horizontal axis. The blue crosses correspond to the non-improved correlators, whereas the orange dots correspond to the improved correlator with $\alpha$ tuned. The plots show that our new correlator \eqref{eq:C_alpha} performs better than the non-improved one, as the values of the orange points are almost always smaller than the blue crosses. However, one could argue that in some cases the value with a reduced number of sources is already good enough even with $\alpha=0$. While this is true, it is also true that this is not known a priori; in every case, the improved correlators give better results regardless, and it is then preferred to the non-improved one.

\begin{figure}[h]
  \centering
  \includegraphics[scale=0.25]{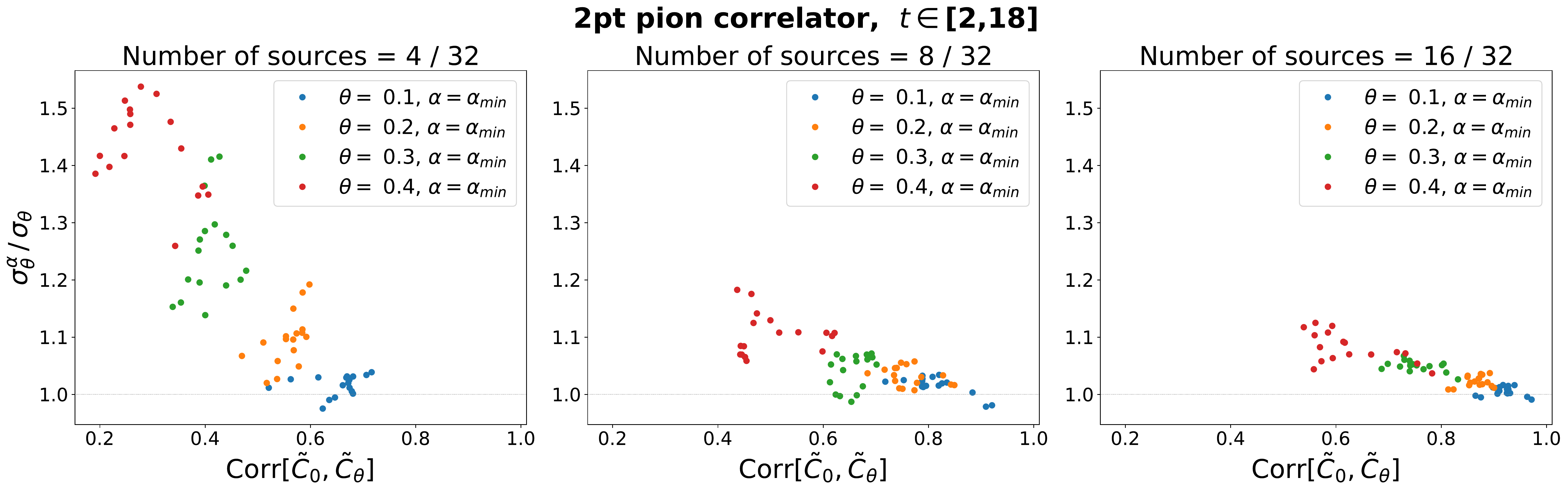}
  \caption{Improvement $\sigma^{\alpha}_\theta / \sigma_{\theta}$ as a function of the correlation 
           $\Corr[\tilde{C}_0,\tilde{C}_{\theta}]$ for every twists.}
  \label{fig:correlation_2pt}
\end{figure}

For what concerns larger twists, we observed that the improvement $\sigma^{\alpha}_\theta / \sigma_{\theta}$ is controlled by the correlation $\Corr[\tilde{C}_0,\tilde{C}_{\theta}]$ between twisted and untwisted correlators, as shown in figure \eqref{fig:correlation_2pt}. The smaller the twist, the higher the correlation and the better is the improvement $\sigma^{\alpha}_\theta / \sigma_{\theta}$. The points in the plot are taken from the range $t\in[2,18]$, which covers the range where one would perform fits.
This observation, in principle, could provide a guidance to estimate the possible gain a priori and could therefore help to plan the computational setup. We see a similar behaviour also in the case of 3pt functions.

\subsection{Form factors}

We now present some results on the $K\rightarrow\pi$ form factors $f_+(q^2)$ and $f_-(q^2)$ defined in terms of the hadronic matrix elements
\begin{align}
    M_{\mu} = \bra{\pi(p_\pi)} V_\mu^{R} \ket{K(p_K)} = f_+(q^2)(p_K + p_\pi)_\mu + f_-(q^2)(p_K - p_\pi)_\mu \, , 
\end{align}
where $V_\mu^{R} = Z_V V_{\mu}$ is the renormalised vector current with $Z_V$ computed from the $\pi\rightarrow\pi$ vector form factor  at $q^2=0$ as, e.g., in \cite{pion_form}.
We can extract the matrix element from the Euclidean time dependence of 2pt correlators $C_{K}(t)$, $C_{\pi}(t)$ and 3pt correlators $C_{\mu,K\pi}(t)$, $C_{\mu,\pi K}(t)$ as
\begin{align}
  M_\mu \overset{t_\pi \gg t \gg t_K}{=} 4Z_V \sqrt{E_K E_\pi} \sqrt{\frac{C_{\mu,K\pi}(t) C_{\mu,\pi K}(t)}{C_K (t) C_\pi (t)}} \, ,
\end{align}
and we can then express the form factors as
\begin{align}
 \begin{dcases}
  f_+(q^2) = \frac{1}{2E_K}\left(M_0 + \frac{E_K-E_\pi}{p_\pi} M_s\right) \, ,\\
  f_-(q^2) = \frac{1}{2E_K}\left(M_0 - \frac{E_K+E_\pi}{p_\pi} M_s\right) \, ,
 \end{dcases}
 \label{eq:form_factors}
\end{align}
%
%
where $M_s$ is the average over the $3$ spatial matrix elements $M_i$, $i=1,2,3$.
We recall that the twist is applied on the light quark in $C_\pi(t)$ and in the 3pt correlators, but not in $C_K(t)$. We then consider each correlator as a function of a different $\alpha_i$
\begin{align}
    M_\mu(\alpha_1, \alpha_2, \alpha_3) \propto  
    \sqrt{\frac{C^{\alpha_1}_{\mu,K\pi}(t) C^{\alpha_2}_{\mu,\pi K}(t)}{C_K (t) C^{\alpha_3}_\pi (t)}} \, ,
 \label{eq:Mmu}
\end{align}
such that the error $\sigma_f^{\alpha}$ of the form factor is a function of $\bm{\alpha}= (\alpha_1, \alpha_2, \dots, \alpha_n)$, i.e. $\sigma_{f}^{\alpha}=\sigma_{f}(\bm{\alpha})$. The values of $\bm{\alpha}$ are then determined by minimizing numerically the error $\sigma_{f}(\bm{\alpha})$, i.e. by imposing
$\nabla_{\bm{\alpha}}\sigma_{f}(\bm{\alpha})=0$. In particular, since the denominator of $M_\mu$ is always the same and the numerator is always different for every values of $\mu$, we have in total $9$ parameters $\alpha_i$. We will refer to this $\bm{\alpha}$ as the \textit{optimized} $\alpha$ (indicated as $\alpha_{opt}$ in the plots), as opposed to the analytical value $\alpha_{min}$ in \eqref{eq:alphamin}.

\begin{figure}[h]
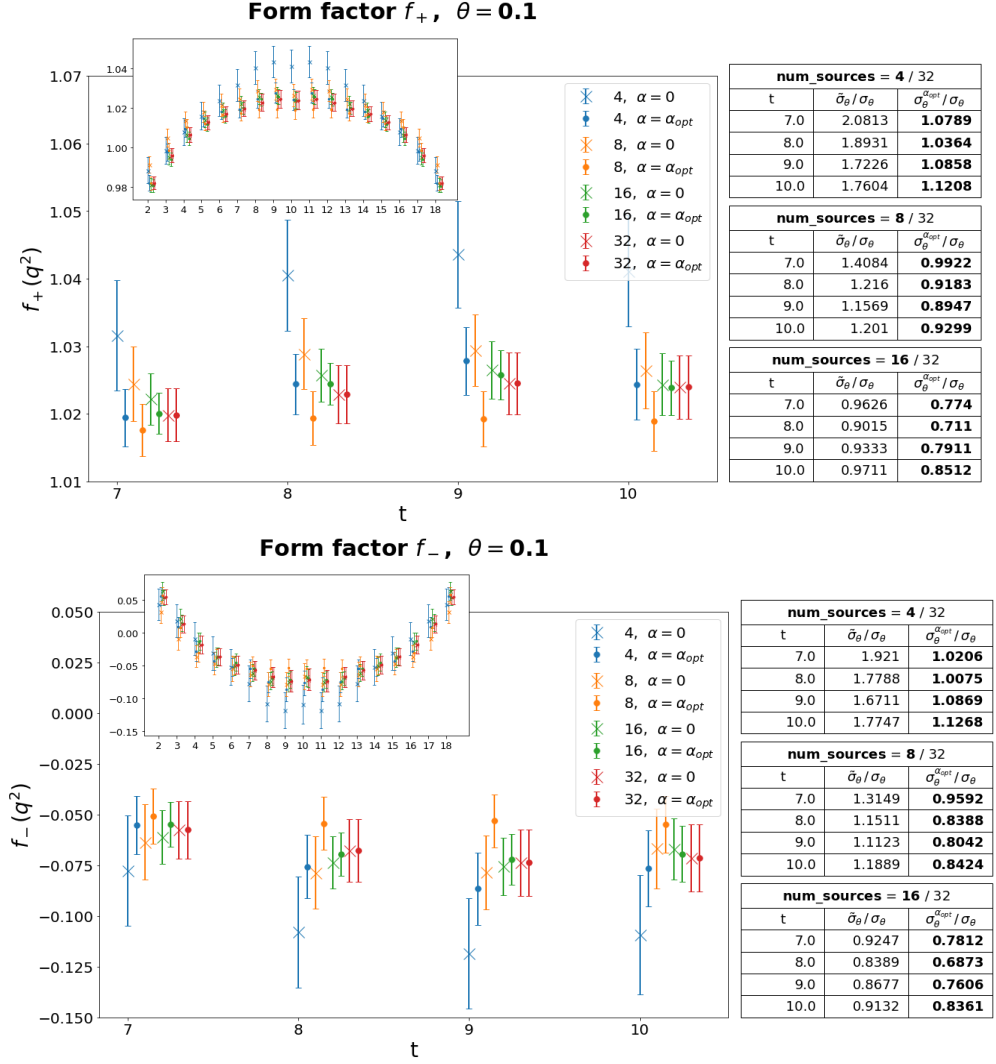

\centering
 \includegraphics[scale=0.28]{plotAB_ALPHA/formp_tw010_alphaopttw000.pdf}
 \includegraphics[scale=0.28]{plotAB_ALPHA/formm_tw010_alphaopttw000.pdf}
 \caption{Form factor $f_+$ and $f_-$ in the range $t\in [7,10]$, where the fit is performed. The tables on the right show the ratio between the error with reduced and full statistics (central column) and the ratio with the optimized technique and the full statistics (right column). }
 \label{fig:f+-}
\end{figure}

In figure \eqref{fig:f+-} we show the results for the $f_+(q^{2})$ and $f_-(q^{2})$ $K\rightarrow\pi$ form factors. In the top left corner, we show the time dependence of the form factors \eqref{eq:form_factors} between source and sink.
In the main plot, we zoom on the subset of points we are using for the final fits. From the plot, it is evident that the optimization technique is successful. The numbers are reported in the table on the right: for every time slice, the central column correspond to the ratio $\tilde{\sigma}_\theta / \sigma_\theta$ and it is compared to the new ratio $\sigma^{\alpha_{opt}}_\theta / \sigma_\theta$. We can clearly see that already with 8 sources out of 32 we get a smaller error than the one with full statistics. In other words, we can achieve a better result at only 1/4 of the computational cost, similarly to what happened with the bare correlators.

\begin{figure}[ht]
\centering
 \includegraphics[scale=0.27]{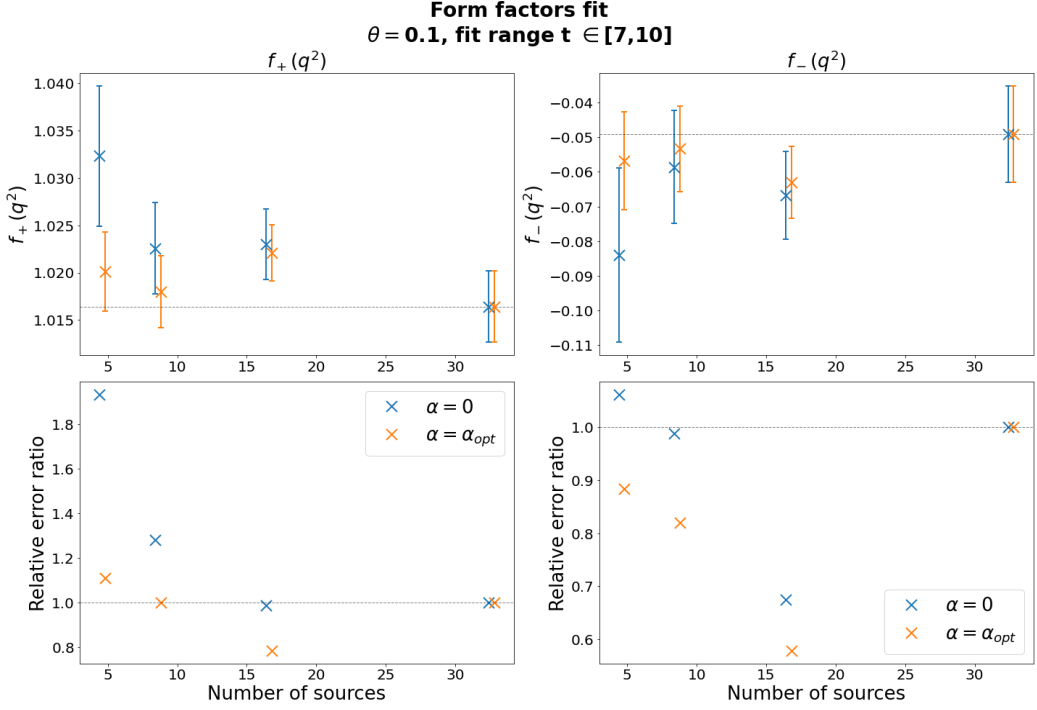}
 \caption{Form factor $f_+$ and $f_-$ fit in the range $t\in [7,10]$. }
 \label{fig:formfit}
\end{figure}

We finally show in figure \eqref{fig:formfit} the values of $f_+$ and $f_-$ obtained from the fit to a constant. In the top plots, we show the values of the form factors as a function of the number of sources, whereas in the bottom plots we report the ratio between relative errors, corresponding to $\tilde{\sigma}_{rel}/\sigma_{rel}$ (blue points) and $\sigma^{\alpha_{opt}}_{rel}/\sigma_{rel}$ (orange points). Again, the plots show that our technique allows us to do better even than the full statistics case for at least $1/4$ of the total number of sources.

\section{P$\pm$A boundary condition}

\begin{figure}[h]
 \centering
 \includegraphics[scale=0.3]{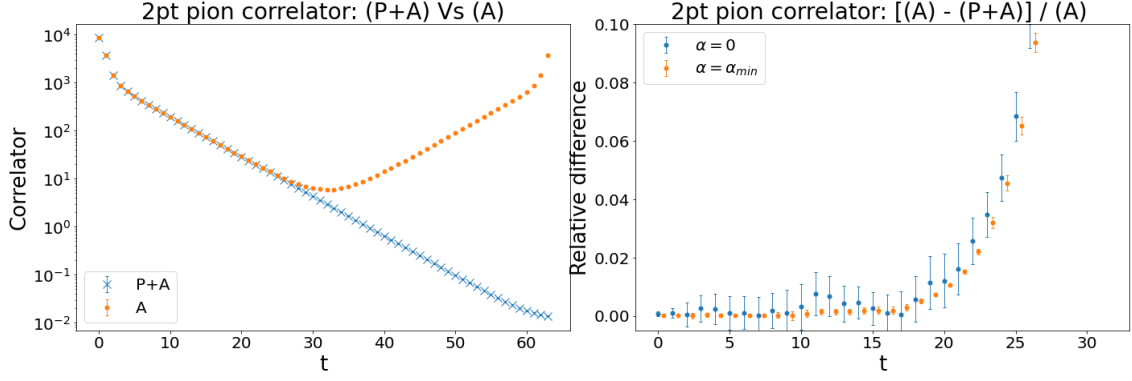}
 \caption{Comparison between correlators with Antiperiodic boundary conditions (A) and Periodic+Antiperiodic (P+A) boundary conditions. On the left, the two correlators are plotted together: (P+A) has period 2T, so only half of it appears. On the right, the relative difference between the two is shown.}
 \label{fig:PA}
\end{figure}

We applied the very same technique also in a different condition, namely the case with Periodic $\pm$ Antiperiodic boundary conditions \cite{PA} in time direction. Given a field $\phi(t)$ with period $T$, we can choose Periodic boundary conditions (P) or Antiperiodic boundary conditions (A) in time direction. For the corresponding fields the following relations will hold
\begin{align*}
 \phi_{P}(t+T) = \phi_{P}(t) \, , \quad \phi_{A}(t+T)= -\phi_{A}(t) \, .
\end{align*}
We can then combine these two fields into $\phi(t)_{P\pm A}= \phi_{P}(t) \pm \phi_{A}(t)$:
both choices of boundary conditions have the effect of doubling the period of the field and have thus the advantage of eliminating the effect of the around-the-world contributions arising from the finite-volume formulation. In figure \eqref{fig:PA} the correlators with Antiperiodic (A) and Periodic+Antiperiodic (P+A) boundary conditions are shown together: we can clearly see that the period of (P+A) is different and that in this case the around-the-world effects are subtracted.
On the right, we plot the relative difference between the two correlators: this shows where the around-the-world effects become significant, which is around $t/a=25$.

Coming now to the new variance reduction technique, we build a new correlator as in \eqref{eq:C_alpha}
\begin{align}
 C_{P\pm A}^{\alpha}(t) = \tilde{C}_{P\pm A}(t) + \alpha(t) ( \tilde{C}_{P/A}(t) - C_{P/A}(t) ) \, ,
\end{align}
where in this case we could use either the Periodic (P) or the Antiperiodic (A) as the full statistics correlator. We will mainly consider $C_{P+A}^{\alpha}(t)$ improved with the Antiperiodic case (A).

\begin{figure}[h]
 \centering
 \includegraphics[scale=0.3]{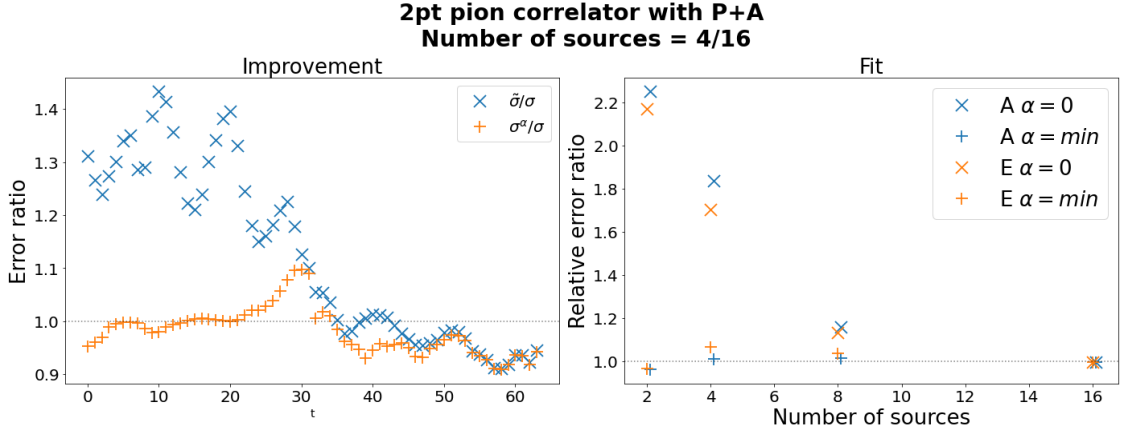}
 \caption{Improvement of the new correlator compared with the reduced statistics correlator (on the right). Relative error ratio of the fit results as a function of the number of sources (on the left).}
 \label{fig:corrPA}
\end{figure}

In figure \eqref{fig:corrPA} we show the results obtained on a 2pt pion correlator with 4 sources out of a total of 16, in analogy with what we did for the twisted case. 
On the left, the plot shows that the ratio 
$\sigma^{\alpha}/\sigma$ built from the improved correlator is much smaller than the ratio $\tilde{\sigma}/\sigma$. The fit on the right is performed using the ansatz \eqref{eq:fit}, and it shows that the new correlator gives the same results as the full statistics correlator for every number of sources. This is not surprising, as the fit range lies typically in the first half of the lattice, where the (A) correlator and (P+A) correlator are almost the same, as shown in figure \eqref{fig:PA}.
Overall, even with these boundary conditions we obtain similar results as in the twisted case.

\section{Summary and outlook}

We have shown in this work that it is possible to obtain precise twisted correlators at a reduced computational cost, starting from the knowledge of the full untwisted correlator. In particular, we have shown that we can define a new twisted correlator \eqref{eq:C_alpha} using a reduced number of sources for the twisted case and exploiting the untwisted correlator; a parameter $\alpha$, determined analytically, is also introduced to minimize the error. The method works well for small twist angles, with a computational cost roughly equal to 1/4 of the total one; the improvement is controlled by the correlation between the untwisted and twisted correlators. A similar analysis can be performed also for the Periodic$\pm$Antiperiodic boundary conditions case. \\
For generic observables, such as the form factors, we can apply the same strategy by tuning a larger number of parameters $\alpha_i$, which can be determined numerically. Even in this case, the technique is giving very promising results, indicating again that we can achieve high precision results at 1/4 of the total cost.

To complete the analysis, one also needs to take into account possible biases introduced in the new correlators.
For the level of statistical errors obtained here, the results seem to suggest that the new constructed correlator leads to unbiased central values, but a complete analytical understanding of the process is required.
Also, the Periodic$\pm$Antiperiodic case needs to be tested for physical calculations, for example the $K-\pi$ scattering \cite{Janowski}, where the control of around-the-world effects plays a crucial role.

\section*{Acknowledgments}

We thank our RBC and UKQCD collaborators for helpful discussions and suggestions.
This work used the DiRAC Extreme Scaling service at the University of Edinburgh, operated by the Edinburgh Parallel Computing Centre on behalf of the STFC DiRAC HPC Facility (www.dirac.ac.uk). This equipment was funded by BEIS capital funding via STFC capital grant ST/R00238X/1 and STFC DiRAC Operations grant ST/R001006/1. DiRAC is part of the National e-Infrastructure.
A.J. is supported by STFC grant ST/T000775/1. A.B. is supported by the Mayflower scholarship in the School of Physics and Astronomy of the University of Southampton.

\end{document}